\documentclass[prb]{revtex4}

\usepackage{bm}

\usepackage{amsmath}
\usepackage{graphicx}

\usepackage[colorlinks=true,linkcolor=blue]{hyperref}
\expandafter\ifx\csname package@font\endcsname\relax\else
 \expandafter\expandafter
 \expandafter\usepackage
 \expandafter\expandafter
 \expandafter{\csname package@font\endcsname}
\fi
\begin{document}

\title{Simulations of {\it in situ} x-ray diffraction from uniaxially-compressed highly-textured polycrystalline targets}

\author{David McGonegle}
\email{d.mcgonegle1@physics.ox.ac.uk.}
\affiliation{Department of Physics, Clarendon Laboratory, University of Oxford, Parks Road, Oxford OX1 3PU, UK}
\author{Despina Milathianaki}
\affiliation{Linac Coherent Light Source, SLAC National Accelerator Laboratory, Menlo Park, CA 94025, USA}
\author{Bruce A Remington}
\affiliation{Lawrence Livermore National Laboratory, Livermore, CA 94550,USA}
\author{Justin S Wark}
\author{Andrew Higginbotham}
\altaffiliation{Now at Department of Physics, University of York, Heslington, York YO10 5DD, UK}
\affiliation{Department of Physics, Clarendon Laboratory, University of Oxford, Parks Road, Oxford OX1 3PU, UK}
\date{\today}

\begin{abstract}
A growing number of shock compression experiments, especially those involving laser compression, are taking advantage of \emph{in situ} x-ray diffraction as a tool to interrogate structure and microstructure evolution.  Although these experiments are becoming increasingly sophisticated, there has been little work on exploiting the textured nature of polycrystalline targets to gain information on sample response.  Here, we describe how to generate simulated x-ray diffraction patterns from materials with an arbitrary texture function subject to a general deformation gradient.  We will present simulations of Debye-Scherrer x-ray diffraction from highly textured polycrystalline targets that have been subjected to uniaxial compression, as may occur under planar shock conditions.  In particular, we study samples with a fibre texture, and find that the azimuthal dependence of the diffraction patterns contains information that, in principle, affords discrimination between a number of similar shock-deformation mechanisms. For certain cases we compare our method with results obtained by taking the Fourier Transform of the atomic positions calculated by classical molecular dynamics simulations.  Illustrative results are presented for the shock-induced  $\alpha$--$\epsilon$ phase transition in iron, the $\alpha$--$\omega$ transition in titanium and deformation due to  twinning in tantalum that is initially preferentially textured along [001] and [011]. The simulations are relevant to experiments that can now be performed using 4th generation light sources, where single-shot x-ray diffraction patterns from crystals compressed via laser-ablation can be obtained on timescales shorter than a phonon period. 

\end{abstract}

\maketitle

\section{Introduction}

The response of matter to shock compression has been a subject of study for more than a half a century \cite{Rice1958,Duvall1963,Davison1979,Swegle1985,Kalantar2005}, and early in this field of research it was recognised that the high transient stresses to which materials could be subject caused them to yield and deform plastically, and, in certain circumstances, to undergo rapid polymorphic phase transitions \cite{Minshall1955,Bancroft1956,Fowles1961,Hayes1974}.  An understanding of the pertinent physics operating at the lattice level that underpins the material response has long been sought, and has been a strong motivator in the development of lattice-level and structural measurement techniques with sufficient temporal resolution to interrogate detailed material response during the deformation process itself.  Of particular relevance to the work presented here are recent advances in ultra-fast x-ray diffraction.  The first diffraction patterns of crystals subjected to shock compression taken during the passage of the shock itself had exposure times of ten of nanoseconds \cite{Johnson1970a,Johnson1971,Johnson1972,Johnson1972a} .  Since this pioneering work significant progress has been made by the use of x-ray sources based on diode-technology \cite{Germer1979}, synchrotron emission\cite{Bilderback2005}, and the use of x-rays emitted from plasmas created by irradiation with high-power lasers \cite{Wark1987,Wark1989}.  More recently however, the development of so-called 4th generation light sources, such as the Linac Coherent Light Source (LCLS), has allowed high-quality single-shot diffraction patterns to be obtained in just a few tens of femtoseconds -- freezing motion on a timescale faster than the shortest phonon period in the system.\cite{Milathianaki2013}

The transient x-ray sources mentioned above have been used to further our understanding of the response of both single-crystals and polycrystalline matter to both shock and quasi-isentropic compression \cite{Johnson1971,Johnson1972a,Wark1989,Whitlock1995,Gupta1999,Rigg1998,D'Almeida2000,Loveridge-Smith2001,Kalantar2003,Kalantar2005,Hawreliak2006,Murphy2010a,Hawreliak2011,Rygg2012,Comley2013,Milathianaki2013}.  A number of x-ray diffraction techniques have been developed to monitor material response, including divergent beam geometry\cite{Kalantar2003a}, white-light Laue diffraction\cite{Suggit2010}, Debye-Scherrer diffraction\cite{Johnson1970a,Johnson1971,Johnson1972,Johnson1972a} and the use of energy-dispersive single-photon counting\cite{Higginbotham2014a}.   While studies of both single-crystals and polycrystalline matter have been undertaken, in the case of polycrystalline samples very little attention has yet been paid to the role of texture in these uniaxial compression experiments -  that is to say the distribution function of grain orientations within a particular polycrystalline specimen.  A number of manufacturing methods, such as rolling and epitaxial growth, result in characteristic textures due to the way in which the materials have been processed. This texture has an influence on a range of physical properties such as strength, electrical conductivity and wave propagation \cite{Wenk2004}. Therefore texture plays an important role in understanding material response.

Furthermore, whilst the texture of a sample will influence its response to rapid uniaxial compression, it will also have a profound influence on the way in which the sample diffracts.   As certain grain orientations are more likely to occur than others, the intensity of a particular Debye-Scherrer ring (corresponding to a certain set of Miller indices) will have a strong azimuthal dependence, and this dependence can in turn be used to extract texture information.   Indeed, significant static studies of the texture of polycrystalline samples have been undertaken with synchrotron sources for many years.  \cite{Wenk1997,Wenk2003,Wenk2004,Wenk2005,Ischia2005,Barton2012,Vogel2012}. Wenk and co-workers provide an overview of the use of synchrotrons in such texture analysis \cite{Wenk2003}. While the texture of a material is often represented by a set of pole figures which can be measured directly via x-ray diffraction, prediction of anisotropic material properties requires knowledge of the full orientational distribution function (ODF). The ODF gives the probability of a crystallite having an given orientation, therefore provides a complete description about the texture of the sample. Pole figures, being a 2D projections of the ODF, result in some texture information being lost, although methods have been developed to obtain an approximate ODF from pole figures \cite{Wenk2003}.

Given that the preferred orientation defined by texture links both the diffraction patterns observed, and the sample response, it is logical to question whether specific information can be gleaned via {\it in situ} diffraction studies of samples with known, well-defined texture. For example, bulk rotations of the crystal lattice, or changes in the crystal structure, such as Martensitic phase changes, will result in an altered texture that could be used to distinguish between different mechanisms of atomic rearrangement. This reorientation has been observed in previous work using both neutron sources \cite{Brown2005} and synchrotrons\cite{Wenk2003,Vogel2012}, although only at relatively modest pressures compared with those we are interested in here.

Ideally one would wish to understand the detailed response to shock compression of samples as a function of their ODF, and to then predict the resultant diffraction patterns.  We outline in section \ref{S:Method} the method by which this could in principle be done.  However, in terms of the results of particular simulations, our present goal is more modest: we restrict ourselves to crystals that are highly fibre-textured -  that is to say all of the grains within the sample initially have very similar orientations with respect to a single axis (the fibre axis), only deviating slightly in angle from that of a high-symmetry direction.  As all of the grains have similar orientations along a given axis, which we also take to be the axis of compression, we might expect that we can, to a reasonable approximation, model certain aspects of their response to shock or quasi-isentropic compression using single-crystal parameters.  In particular, when determining how the material will respond to compression and shear, we assume that the elastic stresses, supported by elastic strains, can be calculated by using elastic constants appropriately chosen to mimic single crystal response along the pertinent directions.  In terms of x-ray diffraction, however, the finite range of orientations determined by the ODF is such that a monochromated, non-divergent incident x-ray beam can diffract from a reasonably large subset of the grains, both in the shocked and un-shocked case.  A similar approach to that outlined above has recently been used to observe, via femtosecond diffraction, the ultimate compressive strength of copper subjected to shock compression at strain-rates of order 10$^9$ s$^{-1}$\cite{Milathianaki2013}. We shall show that breaking the symmetry of the system, such that the direction of the incident x-rays and the target normal (parallel to the compression direction) are no longer antiparallel, allows us to determine specific information about the system under study.

The layout of the paper is as follows.  In section \ref{S:Method} we describe the method by which we determine the analytic diffraction pattern as a function of compression, and the angle that the sample normal makes to the x-ray beam.  The background to the molecular dynamics simulations are outlined in section \ref{S:MD} before, in section \ref{S:Results} providing the results for simulations for the $\alpha$--$\epsilon$ phase transition in iron, the $\alpha$--$\omega$ transition in titanium, and deformation due to  twinning in tantalum.  We then conclude with a discussion of the results, and suggestions for experiments and further work.

\section{Method}\label{S:Method}

We have previously shown how to  calculate the position of Debye-Scherrer rings from a polycrystal strained by an arbitrary deformation gradient in the Voigt limit \cite{Higginbotham2014}, and for the sake of completeness we restate briefly the results here.  We envisage a planar sample orientated such that its surface normal $\bf{n}$ lies along the $z^{\prime}$ direction.  As mentioned above, we allow symmetry to be broken, such that the direction of incidence of x-rays for wavevector $\mathbf{k_0}$ can be non-parallel to $\mathbf{n}$, and along the direction $z$. The rotation matrix $\mathbf{R}$ transforms from the unprimed (x-ray) to primed (target) co-ordinate system.  Debye-Scherrer diffraction occurs when the Laue condition is met: $\mathbf{G}=\mathbf{R}{^T}{\mathbf{G}}^{\prime}=\mathbf{k}-\mathbf{k_0}$, where $\mathbf{k}$ is the wave-vector of the diffracted x-ray.  Deformation of the crystal in real space is represented by the arbitrary deformation gradient $\mathbf{F}$ applied in the target (primed) co-ordinate system, which corresponds to the analogue in reciprocal space of $({\mathbf{F}}^{T})^{-1}$.  With these relations we see that the reciprocal lattice vector for an unstrained crystal in x-ray co-ordinates, $\mathbf{G_0}$, is transformed under deformation to a new reciprocal lattice vector $\mathbf{G}$, where
\begin{align}
\mathbf{G_0} & = \mathbf{R}{^T}\mathbf{F}{^T}\mathbf{RG} = \mathbf{\boldsymbol\alpha G}, \label{eg:transform}
\end{align}
where $\mathbf{\boldsymbol\alpha}  = \mathbf{R}{^T}\mathbf{F}{^T}\mathbf{R}$. By noting that if a plane is to diffract once the sample has been strained, it must meet the Laue condition, and that for the unstrained case, $\left|\mathbf{G_0} \right| = \frac{2\pi}{d_0}$, it is possible to solve equation \ref{eg:transform} to yield the Bragg angle as a function of azimuthal angle around the $z$-axis, $\phi$, for an arbitrary deformation gradient -- that is to say we know the direction of the scattered wavevector, $\mathbf{k}$, and hence can determine via simple ray-tracing where the diffraction will impinge on a distant detector. In Higginbotham \emph{et al.} \cite{Higginbotham2014}, diffraction patterns for numerous deformation gradients and target geometries (with respect to the x-ray beam) are shown.  However, in these cases the sample was assumed to be isotropic in texture, and thus satisfying equation \ref{eg:transform} was deemed a sufficient condition for diffraction to occur.

However, a non-isotropic ODF places further constraints on the possibility of diffraction for a given Bragg angle and azimuthal position around the Debye-Scherrer ring, as the ODF provides the measure of the probability of finding a crystal with a given $\mathbf{G_0}$ existing in the original unstrained sample.  The route forward for simulating diffraction for a crystal with known original ODF under a known arbitrary deformation gradient is now clear: we use the ODF to determine the probability of a given crystallite having the appropriate $\mathbf{G_0}$ existing within the sample, we then determine $\mathbf{G}$ from equation \ref{eg:transform} (and hence $\mathbf{k}$ from the Laue condition), and use ray-tracing to propagate the diffracted beam to the detector assuming that the intensity is proportional to the probability of finding the original $\mathbf{G_0}$, as determined by the ODF, and taking into account factors such as multiplicity.

Although the above approach is general, within the rest of this paper we restrict our study to the particular case of a simple fibre texture where the crystallites have nearly identical orientation in the axial direction, but close to random radial orientation. Our motivations for this are due to the fact that this allows us to treat the mechanical response of the polycrystal to be well-approximated by that of a single crystal with orientation aligned with the fibre axis and that the technique of fibre diffraction under ambient conditions is well established \cite{Polanyi1921,Polanyi1923}.  Furthermore, recent experiments using femtosecond x-rays created with $4^{\textrm{th}}$ generation light sources to diffract from uniaxially compressed samples have employed targets with this type of texture,\cite{Milathianaki2013} and the thin films that have hitherto been used in these experiments often grow with such preferential orientation.

Fibre textured samples have a greatly simplified ODF. For the case of perfect fibre texture, each crystallite has a single crystallographic direction, the fibre orientation, $\mathbf{v_1}$, associated with the reciprocal lattice vector $(h_1 ,k_1 , l_1)$, which for all crystallites are aligned parallel to the sample normal, $\mathbf{n}$. Each grain is then deemed to possess a random orientation when rotated about the axis, $\mathbf{v_1}$.  If we consider a particular plane within a crystallite, with miller indices $(h_2 ,k_2 , l_2)$ (which are the same set of miller indices associated with $\mathbf{G_0}$) to which the reciprocal lattice vector is $\mathbf{v_2}$,  then the value of $\mathbf{\hat v_1} \cdot \mathbf{\hat v_2}$ will be a constant.  However, when $z$ and $z^{\prime}$ are non-parallel, $\mathbf{\hat{v}_1} \cdot \mathbf{\hat{G}_0}$ varies as a function of azimuthal angle around $z$.  Thus, for perfect fibre texture the Debye-Scherrer pattern is not a ring pattern, but an array of points defined by simultaneously satisfying equation \ref{eg:transform}, as well as the condition $\mathbf{\hat{v}_1} \cdot \mathbf{\hat{G}_0} = \mathbf{\hat{v}_1} \cdot \mathbf{\hat v_2}$. Note that this condition does not take into account multiplicity, and therefore one must consider this condition for each member of $\mathbf{v_2}$ in the $\{h_2 ,k_2 , l_2\}$ family, since they do not necessarily result in the same values of $\mathbf{\hat{v}_1} \cdot \mathbf{\hat v_2}$.

However, real fibre-texture samples contain crystallites that are not perfectly aligned axially, and the volume fraction of crystallites with reciprocal lattice vectors $\mathbf{v_1}$, $P(\mathbf{v_1})$, will be a rapidly decreasing function of $\mathbf{\hat v_1} \cdot \mathbf{\hat v_n}$, where now $\mathbf{\hat v_n} = < \mathbf{\hat v_1}>$. We further assume that for a given volume fraction with a certain $\mathbf{v_1}$, the directions $\mathbf{v_2}$ of the normals to the planes with miller indices $[h_2 ,k_2 , l_2]$ within the crystallites will be arranged randomly azimuthally around the axis $\mathbf{v_1}$, such that the possible orientations of $\mathbf{v_2}$ are simply constrained by $\mathbf{v_2} \cdot \mathbf{v_1}$ being equal to the value it would take for a single crystal: that is to say $P(\mathbf{v_2}) \propto P(\mathbf{v_1})$ subject to the $\mathbf{v_2} \cdot \mathbf{v_1}$ constraint.  The diffraction condition is once more defined by simultaneously satisfying equation \ref{eg:transform}, as well as the condition $\mathbf{\hat{v}_1} \cdot \mathbf{\hat{G}_0} = \mathbf{\hat{v}_1} \cdot \mathbf{\hat v_2}$., but now the intensity of the diffraction is proportional to $P(\mathbf{v_1})$.  As the possible directions of $\mathbf{v_1}$ can now vary over a constrained range, our pattern is a series of arcs, rather than points. For the sake of simplicity in our simulations here we assume that the volume fractions of crystallite orientiations are uniform over a small range of angles, i.e. $P(\mathbf{v_1}) = C$, a constant, for $| \arccos (\mathbf{\hat{v}_1} \cdot \mathbf{\hat{G}_0}) - \arccos ( \mathbf{\hat{v}_1} \cdot \mathbf{\hat v_2}) | \le \delta$, and $P(\mathbf{v_1}) = 0$ for $| \arccos (\mathbf{\hat{v}_1} \cdot \mathbf{\hat{G}_0}) - \arccos (\mathbf{\hat{v}_1} \cdot \mathbf{\hat v_2}) | > \delta$.  This method effectively finds the intersection between the Ewald and Polanyi spheres \cite{Polanyi1921,Polanyi1923,Stribeck2009}, however for the case of anisotropic strains, the Polanyi sphere is deformed into an ellipsoid. Note that by changing the x-ray energy or the sample orientation, which varies the size of Ewald sphere or rotates the Polanyi sphere respectively, the position of the arcs on the Debye-Scherrer ring also change, and that by varying these parameters, different parts of reciprocal space can be interrogated.

\section{Molecular Dynamics Simulations}\label{S:MD}
\label{MD_sim}

One advantage of the fibre texture discussed above is that one knows (to within the texture width) the crystallographic orientation of grains with respect to an applied planar compression front.  This is particularly important as crystallite orientation can drastically alter the material response under uniaxial loading \cite{Murphy2010a,Dupont2012,Smith2012,Zong2014}.
In the case that grain size is comparable to sample thickness\cite{Milathianaki2013}, one can  approximate the response of the sample as being close to that of a suitably oriented single crystal.  This is particularly pertinent if one wishes to compare results with those of molecular dynamics, where state of the art polycrystalline simulations are still generally limited to grain sizes of ~5-100\,nm \cite{Bringa2005,Kadau2007,Jarmakani2008,Bringa2010,Gunkelmann2014}, far below the grain sizes of typical experimental samples.\\
In order to relax the requirements on computational power, we present a method of simulating the response of a fibre textured target by manipulation of a single crystal simulation. We do this by first performing a 3D Fourier Transform (FT) of the computed electron density of the single crystal \cite{Kimminau2008}.  This provides us with a momentum space representation of the lattice which describes the allowed scattering vectors for diffraction.\\
Working with the intensity of this FT, we first note that any polar dependence around the compression axis can be neglected due the random rotational distribution of grains in a fibre textured sample.  Considering a cylindrical geometry, we therefore produce a 2D representation of the FT, which flattens the data into its $\left(\rho, z\right)$ components, effectively integrating around $\phi$.  In the case of a perfect (zero texture width) fibre texture, this 2D representation correctly describes all scattering.\\
For the case of finite texture width, one can imagine that the misorientations of the grains are simply related to a rotation about the origin of the 2D representation, and so to mimic the width we sum rotated representations for angles between $\pm\delta$.    This new representation necessarily still retains the cylindrical symmetry required for fibre texture, but correctly accounts for the distribution of grain alignments.  Of course, as $\delta$ becomes larger than a few degrees, the underlying assumption that all grains react in a similar manner to a well aligned single crystal will break down.  However, for this paper we will assume suitably small texture widths of $\approx 5^\circ$ where this approach works well.\\
One can now raytrace simulated diffraction patterns directly from this 2D representation by only considering the $\left(\rho, z\right)$ component of the scattering vectors expressed in this cylindrical target geometry.

\begin{figure}
\begin{center}
\includegraphics[width=0.5\linewidth]{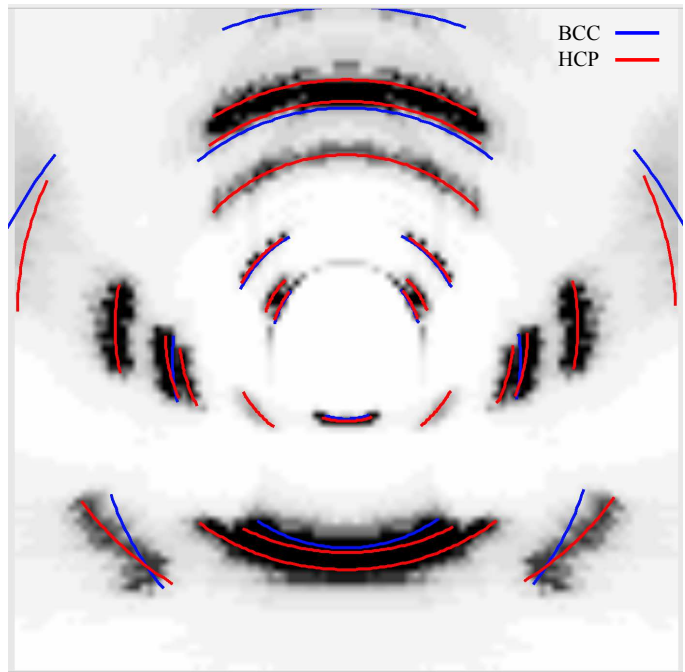}
\caption{(Color online) A simulated ray trace of diffraction from a [001] fibre textured polycrystalline HCP Fe formed under shock compression. The blue overlaid lines show the positions of arcs from uniaxially compressed BCC, while the red lines show the positions of arcs from HCP with OR  $\left[001\right]_{\textrm{bcc}}\left|\right| \left[2\bar{1}\bar{1}0\right]_{\textrm{hcp}}$, $\left[110\right]_{\textrm{bcc}}\left|\right| \left[0002\right]_{\textrm{hcp}}$. For clarity the x-ray energies used to trace the bcc and hcp overlays were offset by 1\%. The corners of the detector are located at a scattering angle $2\mathrm{\theta} = 74.2^{\circ}$.}
\label{fig:Fe_HCP}
\end{center}
\end{figure}

\section{Results}\label{S:Results}

We now present results for three different fibre-textured polycrystals subject to uniaxial compression: iron with a [001] fibre texture, titanium with [0001] fibre-texture, and tantalum with [001] and [011] fibre-textures.  For all three crystals we make assumptions about the deformation mechanisms, and use the approach of section \ref{S:Method} to predict diffraction patterns.  For two of the cases -  iron and tantalum -  we also compare our calculations with diffraction patterns simulated using the method of molecular dynamics, as outlined in section \ref{S:MD}.

\subsection{$\alpha$--$\epsilon$ Phase Transition in [001] Iron}

The $\alpha$--$\epsilon$ phase transition in iron is an example Martensitic transformation, characterised by a collective movement of atoms across distances that are typically smaller than a nearest-neighbour spacing. These type of transitions are well suited to  laser compression studies, since the timescales on which they occur are comparable to the short pulses that can be attained in laser experiments. Importantly, for these non-diffusional transitions, an orientational relationship (OR) exists between the two phases. While the OR does not uniquely determine the mechanism by which the phase transition occurs, it can significantly reduce the number of candidate mechanisms.  The ability to measure the OR \emph{in situ} is therefore highly desirable.

As an example of determination of an OR, we take the case of [001] oriented iron, where the phase transition mechanism is well understood.  Molecular dynamic simulations undertaken by Kadau \emph{et al.} aimed to understand iron's bcc-hcp shock induced phase transition \cite{Kadau2002}.  The results of these investigations were later reproduced with remarkable fidelity in experimental x-ray diffraction studies\cite{Kalantar2005,Hawreliak2006}.  In both MD and experiment, the OR was described as $\left[001\right]_{\textrm{bcc}}\left|\right| \left[2\bar{1}\bar{1}0\right]_{\textrm{hcp}}$, $\left[110\right]_{\textrm{bcc}}\left|\right| \left[0002\right]_{\textrm{hcp}}$ \cite{Kadau2002} .  One can therefore consider this OR as reorienting a fibre textured sample from $\left(002\right)_{\textrm{bcc}}$ to $\left(2\bar{1}\bar{1}0\right)_{\textrm{hcp}}$.

Following Kadau, we simulate a 100x100x800 cell (288x288x2301\AA)  iron single crystal shocked along the [001] direction by 0.7\,km $\textrm{s}^{-1}$  piston using the same Voter-Chen potential used in Kadau's work \cite{Kadau2002}. For this piston velocity, the material does not reach the 18.4\% uniaxial compression needed to create ideal HCP, instead reaching only 13.8\% \cite{Hawreliak2006}, resulting in an anisotropically strained HCP structure. A 3D FT was performed on a section of the material behind the shock front. The FT was modified to mimic that of a fibre textured polycrystal, using the method described in section \ref{MD_sim}. Figure \ref{fig:Fe_HCP} shows the resultant ray trace \cite{Kimminau2008} for a detector in transmission geometry, using a 12\,keV x-ray source and with the sample normal rotated at an angle $30^{\circ}$ to the incoming x-rays.  The overlaid red lines show the predicted diffraction pattern (using the methods of section \ref{S:Method}) of strained HCP iron described above, with a c/a ratio of 1.73, a texture direction along $\left[11\bar{2}0\right]_{\textrm{hcp}}$ and a textured width of $5^{\circ}$. The blue lines show the pattern from 13.8\% uniaxially compressed BCC. As expected, there is agreement between the raytrace and predictions from the molecular dynamics simulations, supporting the validity of the approach outlined in section \ref{S:Method}.\\ 

\begin{figure}[h!]
\begin{center}
\includegraphics[width=0.5\linewidth]{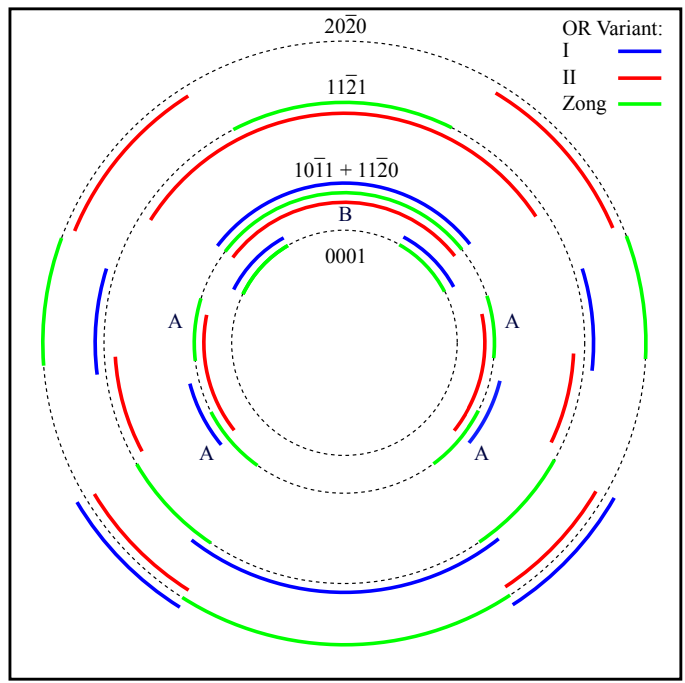}
\caption{(Color online) The predicted diffraction pattern  using 7.5\,keV x-rays  from the $\omega$ phase of a shocked $\left[0001\right]$ fibre textured Ti sample,  with the sample normal rotated at an angle $25^{\circ}$ to the incoming x-rays. The green arcs labelled A and B correspond to the $\left\{10\bar{1}1\right\}$ and $\left\{11\bar{2}0\right\}$ sets of planes respectively. The corners of the diagram are located at a scattering angle $2\mathrm{\theta} = 70.5^{\circ}$.}
\label{fig:Ti}
\end{center}
\end{figure}

\subsection{$\alpha$--$\omega$ Phase Transition in [0001] Titanium}

\begin{table}[tdp]
\caption{The three ORs that correspond to various proposed mechanisms of the $\alpha - \omega$ phase transition in [0001] Titanium}
\begin{center}
\begin{tabular}{ccc}
Variant & Orientational Relationship & Ref\\
\hline
I & $\left(0001\right)_{\alpha}\left|\right| \left(10\bar{1}1\right)_{\omega}$  , $\left[10\bar{1}0\right]_{\alpha}\left|\right| \left[\bar{1}011\right]_{\omega}$ & \cite{Usikov1973,Trinkle2003,Song1995} \\
II & $\left(0001\right)_{\alpha}\left|\right| \left(1\bar{2}10\right)_{\omega}$  , $\left[1\bar{2}10\right]_{\alpha}\left|\right| \left[0001\right]_{\omega}$ & \cite{Usikov1973,Silcock1958}\\
Zong & $ \left(0001\right)_{\alpha}\left|\right| \left(10\bar{1}0\right)_{\omega}$  , $\left[10\bar{1}0\right]_{\alpha}\left|\right| \left[11\bar{2}3\right]_{\omega}$ & \cite{Zong2014}
\end{tabular}
\end{center}
\label{tab:TiORs}
\end{table}

The group IV transition metals titanium (Ti), zirconium (Zr) and hafnium (Hf) have a hexagonal close packed structure ($\alpha$) under ambient conditions, but exhibit a Martensitic phase transition to another hexagonal structure ($\omega$) under high pressure. Although the $\alpha$--$\omega$ transition is well established, the mechanism by which it occurs, and therefore the OR between the phases, is still not fully understood, especially under different loading conditions. A summary of the ORs for various proposed mechanisms is given in table \ref{tab:TiORs}. The first two ORs were proposed by Usikov and Zilbershtein \cite{Usikov1973} in a TEM study of statically compressed Zr and Ti, by arguing that the transition occurs via an intermediate $\beta$ phase, and are usually referred to as Variant I and II. Earlier work by Silcock \cite{Silcock1958} on the $\omega$ phase in Ti alloys proposed a different direct mechanism which corresponds to the Variant II OR. Later computational work by Trinkle \cite{Trinkle2003} demonstrated that a new mechanism, known as TAO-1, had a lower energy barrier than that proposed by Silcock. This new mechanism produced the Variant I OR. Experimental work by Song and Gray \cite{Song1995} observed the OR $\left(0001\right)_{\alpha}\left|\right| \left(10\bar{1}1\right)_{\omega}$, $\left[10\bar{1}0\right]_{\alpha}\left|\right| \left[11\bar{2}3\right]_{\omega}$, although independent analysis re-examining this data led to the conclusions that is actually a subset of Variant I\cite{Jyoti1997}. We have therefore associated Song and Gray's work with Variant I in table \ref{tab:TiORs}. 

Molecular dynamics simulations performed by Zong \emph{et al.} \cite{Zong2014} found that Ti shocked along the $\left[0001\right]$ direction resulted in a mechanism that gave an OR differing from both Variant I and II, as listed in the table. Note that each OR results in a different fibre texture direction for the $\omega$-phase for uniaxial compression of an initially fibre-textured $\alpha$-phase target, and thus a target with the normal tilted with respect to the incident x-ray beam is well suited to provide some discrimination between variants, therefore providing some insight into possible transformation mechanisms.

The work by Zong \cite{Zong2014} found that Ti shocked along the $\left[0001\right]$ direction by a  piston with a velocity of 0.9\,km $\textrm{s}^{-1}$ resulted in transformation to the $\omega$ phase, with lattice parameters $a_{\omega} = 4.61\,\textrm{\AA}$ and $c_{\omega} = 2.82\,\textrm{\AA}$. Using these values, we calculate the diffraction pattern from the $\omega$ phase of a shock-compressed $\left[0001\right]$ sample of Ti with an angular texture width of $\pm 5^{\circ}$.  The sample normal is set at angle $25^{\circ}$ to the incoming x-rays, which have an energy of at 7.5\,keV.  The predicted diffraction patterns for each of the possible variants are shown in Figure \ref{fig:Ti}.  The blue, red and green lines correspond to the diffraction from $\omega$ material of Variant I, Variant II and the Zong OR respectively, while the dotted black line corresponds to the diffraction from an untextured polycrystalline sample. For clarity, the blue lines have been slightly offset outside the true Debye-Scherrer rings, while the red lines have been slightly offset inside. A clear difference can be seen in the diffraction patterns for the different variants, thus allowing the OR and hence a subset of mechanisms to be determined by the azimuthal position of the diffracting arcs around the Debye-Scherrer ring.  It is important to note that the ability to discriminate clearly between all three ORs is not guaranteed, and relies on a judicious choice of both x-ray energy and tilt angle.   

Figure \ref{fig:Ti} also demonstrates that lines with similar $d$-spacings do not necessarily appear at similar azimuthal positions. For example in an ideal $\omega$ crystal, the $\left\{10\bar{1}1\right\}$ and $\left\{11\bar{2}0\right\}$ planes have the same spacing and are therefore completely unresolvable by powder diffraction from an untextured sample. However, since these planes form different angles to the sample normal, within a textured sample their corresponding arcs appear at different azimuthal angles around the Debye-Scherrer ring, allowing them to be resolved. This is shown in Figure \ref{fig:Ti}, where the green arcs labelled A correspond to the $\left\{10\bar{1}1\right\}$ set of planes, while the green arc labelled B corresponds to the $\left\{11\bar{2}0\right\}$ set of planes. By resolving these two arcs, we are able to gain information that cannot be obtained via powder diffraction from an untextured sample; in this case on any small departure from the ideal c/a ratio. We note that it is only possible to resolve lines with similar $d$-spacings if the angle between G-vectors of each of the planes and the sample normal is significantly different. 

\begin{figure*}
\begin{center}
\includegraphics[width=1\textwidth]{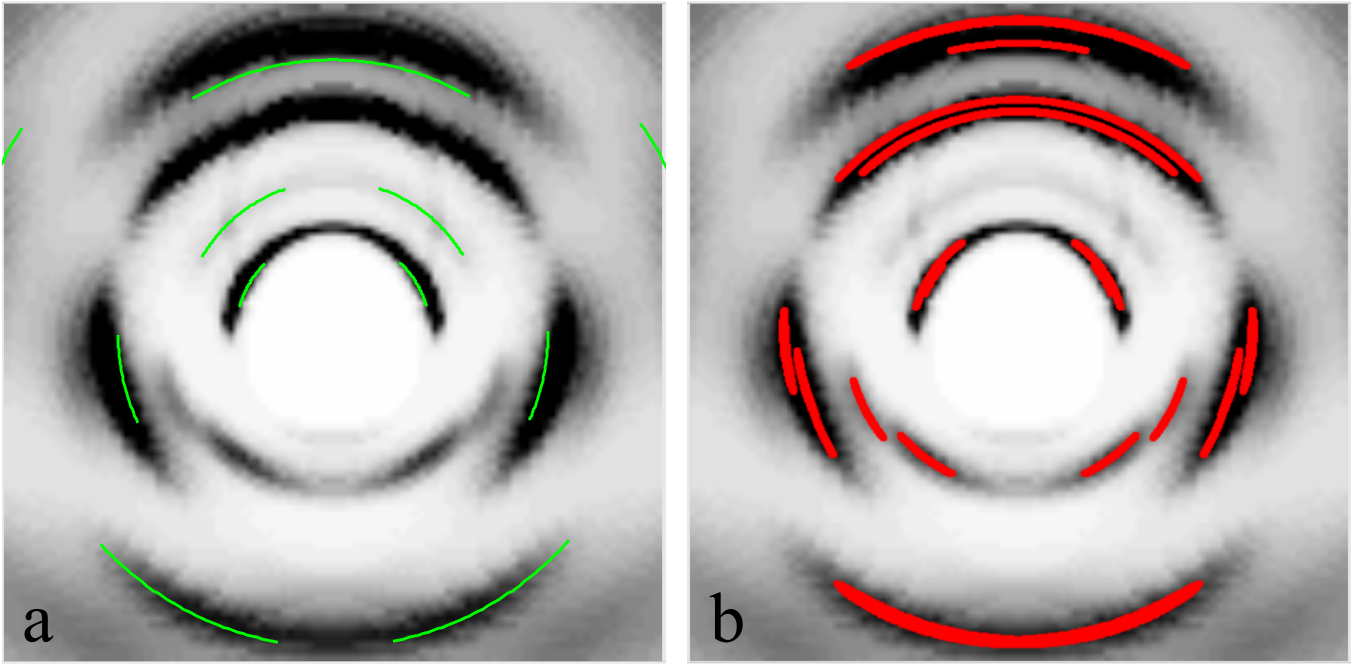}
\caption{(Color online) Simulated diffraction from  a [001] fibre textured polycrystalline Ta under shock compression. Overlaid are the predicted diffraction patterns for: (a). the untwinned case, with elastic strains of  $\epsilon_{t} = -0.029, \epsilon_{l} = -0.136$ shown in green and (b) the twinned case, with elastic strains of $\epsilon_{t} = -0.052, \epsilon_{l} = -0.109$ , shown in red. The corners of the detector are located at a scattering angle $2\mathrm{\theta} = 66.3^{\circ}$.}
\label{fig:Ta_001_Raytrace}
\end{center}
\end{figure*}

\subsection{Twinning in [001] and [011] Tantalum}

Tantalum provides an appealing case to study, owing to its multitude of competing plasticity mechanisms (a combination of dislocation and deformation twinning mediated responses). This is of particular interest, as Debye-Scherrer diffraction in polycrystalline samples with completely random texture cannot distinguish between slip and twinning. Most experimental work into twinning under uniaxial dynamic compression has been performed using either explosive lenses or gas guns\cite{Murr1997,Hsiung2000,Florando2013,Pang2014}. However, to date, time resolved laser diffraction experiments have failed to yield any evidence of twinning \emph{in situ}, although residual twinning has been observed in laser driven shock recovery experiments\cite{Lu2012,Lu2013}.

As with phase transitions, the reorientation of the lattice caused by twinning will result in a different crystallographic direction being oriented along the fibre direction. This may, in turn, lead to a signature in the diffraction pattern. A similar technique using neutron diffraction has been used to observe twinning in magnesium, which occurs at much lower pressures than in tantalum \cite{Brown2005}.

Molecular dynamics simulations performed by Higginbotham \emph{et al.} have predicted a significant twinning fraction in [001] tantalum under shock compression\cite{Higginbotham2013b}. In that work,  the sample was found to be almost completely twinned when compressed by a piston with a velocity of 0.9 km $\textrm{s}^{-1}$, corresponding to a uniaxial compression in the elastic wave by 18\%. The authors noted that after an initial uniaxial compression of 18.4\%, twinning could be achieved by shuffling alternating $\left(\bar{1}12\right)$ planes in the $\left<111\right>$ direction. They therefore proposed this to be the mechanism by which the twinning occurred, with the material reaching its final state via elastic relaxation, although they caution that, given the relatively simple nature of the potential used, they do not claim to exactly model what will occur in practice in shocked Ta. However, the observed shuffling provides a possible mechanism for how twinning of bcc materials may occur under shock compression.

\begin{figure*}
\begin{center}
\includegraphics[width=1\textwidth]{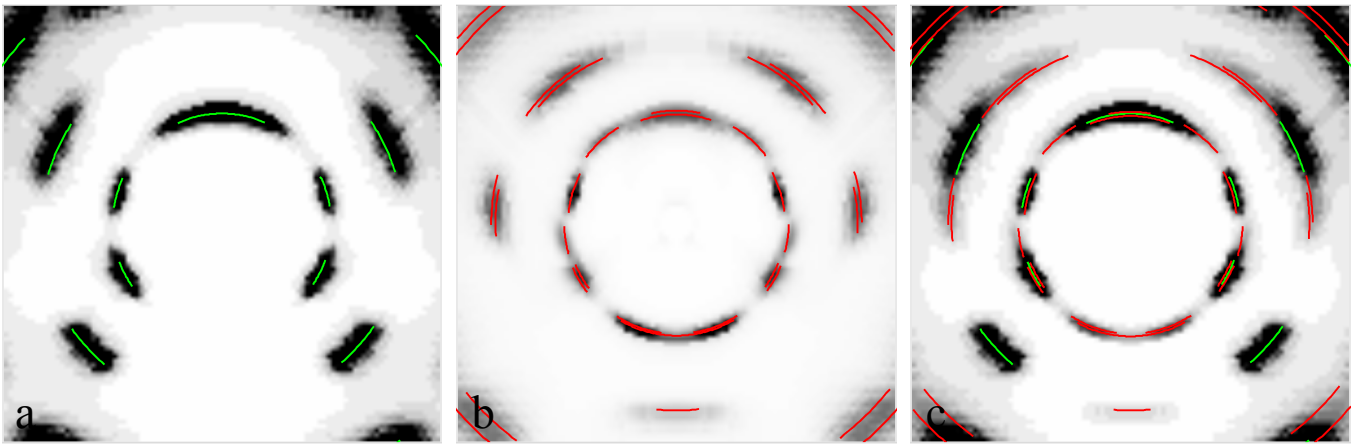}
\caption{(Color online) Simulated diffraction from  a [011] fibre textured polycrystalline Ta under shock compression. Overlaid are the predicted diffraction patterns for: (a). the untwinned case, with elastic strains of  $\epsilon_{t} = -0.045, \epsilon_{l} = -0.075$, shown in green, (b) the twinned case, with the $[\bar{9}22]$ direction close to the compression axis, shown in red, and (c) both the untwinned and twinned case together. The corners of the detector are located at a scattering angle $2\mathrm{\theta} = 51.3^{\circ}$.}
\label{fig:Ta_011_Raytrace}
\end{center}
\end{figure*}

We repeated the simulations in this work, using a 100x100x800 cell (330x330x2644\AA)  Ta single crystal, modelled using the EFS potential \cite{Dai2006}, and shocked along the $\left[001\right]$ direction by a piston travelling at 0.9\,km $\textrm{s}^{-1}$. The per atom structure factor (PASF) \cite{Higginbotham2013b} was used to distinguish between twinned and untwinned material in the plastically deformed material behind the shock front. A 3D FT was performed on a stable region within the plastically deformed material behind the shock front, as well as the twinned region separately, and both were modified in the way described in section \ref{MD_sim}. The diffraction pattern was simulated assuming a 12\,keV x-ray source, with the angle between the incoming x-rays and the sample normal being $25^{\circ}$.  The results are shown in Figure \ref{fig:Ta_001_Raytrace}. For the case of slip, while small rotations of the crystal lattice have been observed \cite{Suggit2012}, no large reorientation is expected, and thus the fibre orientation will remain close to $\left[001\right]$. However, by comparing the observed and predicted positions of the diffraction arcs for a $\left[001\right]$ textured sample including the strains described above (shown in Figure \ref{fig:Ta_001_Raytrace}a), it is clear that there has been significant reorientation of crystallites within the sample, indicative of twinning.

In Figure \ref{fig:Ta_001_Raytrace}b we plot the predicted diffraction pattern using the methods of section \ref{S:Method} assuming that the sample has undergone twinning, and is subject to the longitudinal and transverse strains noted above. In order to find the diffraction pattern resulting from the shuffling mechanism above, one must consider how this affects the lattice vectors of the crystal. The sample is first compressed uniaxially along the z axis by $18.4\%$. The shuffling then has the effect of causing the crystal to be reflected in the $\left(112\right)$ twinning plane. Note that as crystal is uniaxially compressed, this causes the $\left[001\right]$ direction to be reflected to the $\left[111\right]$ direction of the compressed crystal, rather than the $\left[221\right]$ direction expected under hydrostatic conditions (see Figure 6 of Higginbotham \emph{et al.}\cite{Higginbotham2013b}). Lastly, the twinned material returns towards the hydrostat, by relaxing along the longitudinal direction and compression along the transverse directions in the new rotated coordinates. The final longitudinal and transverse strains were measured to be -0.109 and -0.052 respectively. However, in the textured sample, the effect of a finite texture width must also be considered. In this case, the initial compression occurs at slight angle to the texture direction, which results in a slightly different orientation of the twinning plane. The crystal is then reflected in this altered twinning plane, and relaxes as before. To create the predicted diffraction pattern, the lattice vectors of many crystallite orientations within the desired texture width were deformed by the method given above. These were then used to calculate deformed reciprocal lattice vectors, and thus the resultant diffraction pattern.

Excellent agreement can be seen between the analytic solution, and the MD simulation, demonstrating that twinning has occurred, although slight differences can be observed which are due to the small angle assumptions used in section \ref{MD_sim} to simulate the 3D FT of a fibre textured target. Additionally, there are some very weak arcs in the data corresponding to plastically compressed material (Figure \ref{fig:Ta_001_Raytrace}a). The ratio of intensities of lines from twin and slip deformed material is indicative of the twin fraction.

While most theoretical work on twinning in Ta has concentrated on the $\left[001\right]$ direction, recent MD studies by Ravelo \emph{et al.} have suggested shocking along the $\left[011\right]$ direction may be more favourable for deformation twinning, due to the lower observed shear stress threshold for twin nucleation\cite{Ravelo2013}. This agrees with gas gun recovery work on Ta single crystals, which found a significantly higher twin volume fraction in shocks along $\left[011\right]$, compared to the $\left[001\right]$ and $\left[111\right]$ directions\cite{Florando2013}. Furthermore, as the $\left[011\right]$ direction is the preferred direction for epitaxial growth of fibre textured thin films, this direction is particularly well suited for this technique. The $\left[011\right]$ orientation exhibits two types of $\left\{112\right\}\left<111\right>$ twin systems, which result in different fibre orientations. Under hydrostatic conditions, the first type causes no change in the fibre texture, while the second causes a reorientation to the $\left[\bar{4}11\right]$ direction. However, only the second type has non-zero Schmid factors, and it is therefore assumed that only twins of this type occur. It follows that diffraction arcs corresponding to a $\left[001\right]$ fibre orientation are from untwinned material, while arcs corresponding to fibre orientations close to the $\left[\bar{4}11\right]$ type directions are from the two twinned variants.

We repeated the simulations in this work, using a 100x100x800 cell (330x330x2644\AA) Ta single crystal, single crystal, modelled using Ravelo's Ta1 EAM potential \cite{Ravelo2013}, and shocked along the $\left[011\right]$ direction by a piston travelling at 0.62\,km $\textrm{s}^{-1}$. Again the PASF was used to distinguish between twinned and untwinned material, and a 3D FT was performed on each region separately. In the twinned material, the $\left[\bar{9}22\right]$ direction of the compressed crystal is close to the compression axis, which is consistent with twinning after an initial uniaxial compression, similar to the [001] case. The diffraction pattern was simulated assuming a 10\,keV x-ray source, with the angle between the incoming x-rays and the sample normal being $45^{\circ}$. The results are shown in Figure \ref{fig:Ta_011_Raytrace}. Since the twinning mechanism in [011] Ta is not well understood, the predicted pattern was produced for the structure found measured with the FT.   Again, good agreement is seen between the analytic solution and the MD simulation, although in this case, there are strong arcs corresponding to both twinned and untwinned material, suggesting a significant amount of both are present in the sample.

\section{Discussion}

The examples we have given above demonstrate that x-ray diffraction from uniaxially compressed fibre-textured targets can in principle yield information on deformation mechanisms, be they due to phase transformations or twinning.  The breaking of the symmetry of the problem, by tilting the normal of the  polycrystalline target with respect to the incident x-ray beam allows the encoding of such information in the azimuthal distribution of intensity in the Debye-Scherrer rings.  We envisage that the methods that we have outlined here will aid in the design of experiments that have as their goal the elucidation of such mechanisms.  It is worth considering, however, that the choice of initial fibre direction is important in determining what structural information can be extracted. In particular it should be noted that for [001] fibre oriented Fe and Ta samples, these orientation do not have the  lowest surface energy, and thus are not the typical orientations in which thin polycrystalline foils of these materials grow. It may thus be that some effort is required to fabricate suitable samples.  This is not an issue for the case of [0001] Ti or [011] Ta, which are usually grown with these textures. Beyond the four demonstration cases given above, it is clear that further work could concentrate on a variety of different samples and deformation mechanisms.  In addition, we believe that the technique may have other advantages for the study of samples subject to shock or quasi-isentropic compression.  Owing to the high strain rates present in such experiments, high dislocation densities \cite{Bringa2006} or small grain sizes under phase transformation may ensue \cite{Kadau2002,Hawreliak2006}, resulting in broad diffraction peaks that are hard to resolve simply in terms of scattering angle, and thus would not necessarily be easily amenable to study by techniques such as Rietveld refinement.  However, tilting of a target and separation of diffraction peaks azimuthally offers a possible route to finding structural solutions under the extreme pressures that can be obtained via laser-ablation.  We believe that the technique we have outlined could have application to laser based shock and compression experiments that make use of emergent $4^{\textrm{th}}$ generation light sources, which offer incredibly bright, narrow bandwidth x-ray sources, with unprecedented temporal resolution.

\section{Ackowledgments}
DM acknowledges support from LLNL under Subcontract No. B595954.  AH acknowledges financial support from AWE.

\end{document}